\documentclass[oneside,a4paper,english,links]{alioeuro}
\usepackage{graphicx}
\usepackage{amsmath,amsfonts}
\usepackage{mathrsfs}
\usepackage{algorithm}
\usepackage{pstricks}

\title{Solving the 0--1 Multidimensional Knapsack Problem with \\Resolution Search}

\author[a]{Sylvain Boussier}
\author[a]{Michel Vasquez}
\author[a]{Yannick Vimont}
\author[b]{Sa\"id Hanafi}
\author[c]{Philippe Michelon}
\affil[a]{LGI2P, Ecole des Mines d'Al\`es, Parc scientifique Geroges Besse, 30035 N\^imes, France,
\texttt{\{Sylvain.boussier,Michel.Vasquez,Yannick.Vimont\}@ema.fr}}

\affil[b]{LAMIH, Universit\'e de Valenciennes et du Hainaut-Cambr\'esis Le Mont Houy - BP 311, 59304 Valenciennes, France, \texttt{Said.Hanafi@univ-valenciennes.fr}}

\affil[c]{LIA, Universit\'e d'Avignon et des Pays de Vaucluse, 339 chemin des Meinajaries, BP 1228, 84911 Avignon Cedex 9, France, \texttt{Philippe.Michelon@univ-avignon.fr}}

\begin{document}
\vspace{3cm}

\maketitle



\begin{abstract}
We propose an exact method which combines the resolution search and branch \& bound algorithms for solving the 0--1 Multidimensional Knapsack Problem. This algorithm is able to prove large--scale strong correlated instances.  The optimal values of the $10$ constraint, $500$ variable instances of the \texttt{OR-Library} are exposed. These values were previously unknown.
\end{abstract}

\section{Introduction}

In this article, we present a new exact method which hybridizes the resolution search of \cite{Chvatal1997} and a branch \& bound algorithm inspired by a previous work from \cite{Vimont2008} for the 0--1 Multidimensional Knapsack Problem (01MKP). The 01MKP is a well--known opitmization problem which can be viewed as a resource allocation model and can be stated as follows:
\begin{eqnarray}
(P)\hspace{0.5cm} \text{Maximize\hspace{0.5cm}} & {\displaystyle \sum_{j=1}^{n} c_j x_j}& \label{obj}\\
\text{subject to\hspace{0.5cm}}  & {\displaystyle \sum_{j=1}^{n}a_{ij}x_j \leq b_i} & \hspace{0.5cm}i=1,...,m \label{ct1}\\
&{\displaystyle x_j\in \{0,1\}} & \hspace{0.5cm}j=1,...,n \label{ct2}
\end{eqnarray}
where $n$ is the number of items, $m$ is the number of knapsack constraints with capacities $b_i$ ($i=1,2,...,m$), $c \in {\mathbb{N}}^n$, $A \in {\mathbb{N}}^{m \times n}$ and $b \in {\mathbb{N}}^m$. Each item $j$ ($j=1,2,...,n$) yields $c_j$ units of profit  and consumes a given amount of resource $a_{ij}$ for each knapsack $i$.

The proposed approach is centered on two main 01MKP results: (i) the consideration of a reduced costs constraint  based on the reduced costs at the optimality  of the problem's LP--relaxation (see \cite{Balas1980,Oliva2001}) and (ii) the decomposition of the search space in several hyperplane where the number of items to choose is fixed at a given integer value (see \cite{Vasquez2001,Vimont2008}). Our algorithm is self--sufficient and does not require any lower bound as starting value. We show that the structure of resolution search enables to explore partially and iteratively different subproblems (\emph{hyperplanes with a fixed number of items}) while keeping completeness. This way of exploration enhances the diversification of the search and permits to improve the lower bound rapidly.  For each hyperplane, this lower bound, associated with  the upper bound (\emph{tighter than the classical one given by the LP--relaxation}), enforces the strength of the so--called \textsl{reduced costs constraint} widely used in the algorithm. Roughly speaking, if we consider that the variables are sorted in decreasing order of their reduced costs (\emph{basic variables at the bottom of the list}), the enumeration of the first variables is carried out by resolution search where the enumeration of the remaining variables is tackled by our specific branch \& bound. Our approach proved all the  optimality of all the $10$ constraint, $500$ variable instances. These optimal solutions were previously unknown.

\section{Resolution search}
\label{rs}

Resolution search was proposed by \cite{Chvatal1997} as an alternative to branch \& bound for mixed 0--1 linear programming problems. This approach, based on an original exploration of the search space, uses the information brought by the fails that have occurred during the search to progressively shrink the search tree. Each time a terminal node is reached, a minimal partial instantiation of the variables responsible for the fail is identified. This partial instantiation, which corresponds also to a terminal node, is then recorded in a specific way in order to discard the corresponding subtree from the search space and to provide the next node in the exploration. The specificity of the recording mechanism allows the algorithm to preserve space memory while keeping completeness.

Globally, resolution search is composed of two main elements: (i) a set of partial instantiations (denoted path--like family $\mathscr{F}$) corresponding to fails encountered during the search and recorded as boolean clauses and (ii) an function called \texttt{obstacle} which uses the information brought by the family $\mathscr{F}$ to explore promising parts of the search space.  From a partial instantiation $u(\mathscr{F})$ derived from $\mathscr{F}$, which is not an already explored instantiation, \texttt{obstacle} performs two different phases:
\begin{enumerate}
\item \textsl{The waxing phase} which replaces step by step the free components $u(\mathscr{F})_j = *$ by $0$ or $1$ until a terminal node  (denoted $u^*$) is reached (a terminal node is a partial instantiation of the variables which does not lead to any improvement of the best known lower bound).
\item \textsl{The waning phase} which tries to find a minimal element $S$ of  $\left\{ {0,1,\ast } \right\}^n$ such that $S \sqsubseteq u^*$ and $S$ is a terminal node (formally, let $u$ and $v$ be two vectors of $\left\{ {0,1,\ast } \right\}^n$, we call $u$ an \textsl{extension} of $v$ and we write $u \sqsubseteq v$,  if  $v_j = u_j \text{ whenever } u_j \neq \ast$).
\end{enumerate}
Once the minimal element $S$ is identified, it is recorded in a specific way as a clause in $\mathscr{F}$. The structure of $\mathscr{F}$ enables both to guide the search toward prosimising parts of the search space and to guaranteed the completeness using the resolution / refutation principle. Indeed, if the minimal element $S$ correspond to  the instantiation $(\ast,\ast,...,\ast)$ then the best known solution is an optimal soltion of $P$.  The algorithm \ref{alg1}. represents the function \texttt{Resolution\_Search} which  takes as parameter a lower bound (LB) of the problem.
\begin{figure}[!h]
\caption{Resolution Search Algorithm}\label{alg1}
\begin{footnotesize}
\begin{tabular}{l}
\texttt{Resolution\_Search(LB)}\\
\{\\
\hspace{0.5cm}$\mathscr{F} = \emptyset$;\\
\hspace{0.5cm}\texttt{u = (*, *, ..., *);}\\
\hspace{0.5cm}\texttt{while(1)} \{\\
\hspace{1cm}\texttt{try = obstacle(}u,\texttt{LB},$S$);\\
\hspace{1cm}\texttt{if(try > LB)}\hspace{0.3cm}\texttt{LB = try;}\\
\hspace{1cm}\texttt{add} $S$ \texttt{to} $\mathscr{F}$ and update $\mathscr{F}$;\\
\hspace{1cm}u = $u(\mathscr{F})$;\\
\hspace{1cm}\texttt{if($(*, *, ..., *) \in \mathscr{F}$)}\hspace{0.3cm} \texttt{Break;} \\
\hspace{0.5cm}\}\\
\}
\end{tabular}
\end{footnotesize}
\end{figure}

\section{General approach}

\label{genapp}
The 01MKP is tackled by decomposing the problem in several subproblems where the number of items to choose is fixed at a given integer value. Considering that $LB$ is a lower bound of the problem, we can define   $X=\{x \mid A \cdot x \leq b,  c \cdot x \geq LB+1, x \in \{0,1\}^n\}$ as the set of feasible solutions strictly better than $LB$ and consider the following problems:
$$P^+: \text{ maximize } \{1\cdot x  \mid  x \in \bar X\} \hspace{0.5cm} \text{ and } \hspace{0.5cm} P^-: \text{ minimize } \{1\cdot x  \mid  x \in \bar X\}$$
where $1$ is the vector of dimension $n$ with all its components equal to one and  $\bar X=\{x \mid A \cdot x \leq b, c \cdot x \geq LB+1, x \in [0,1]^n\}$. In what follows, we note $v(P)$ the optimal value of a given problem $P$. Let $k_{min} = \left\lceil {v(P^-)}\right\rceil$ and $k_{max} = \left\lfloor {v(P^+)}\right\rfloor$, then we have $v(P) = max\{v(P_k) \mid k_{min} \leq k \leq k_{max}\}$ where
$$P_k: \text{ maximize } \{cx \mid x \in X, 1\cdot x = k\}.$$
Solving the 01MKP by tackling separately each of the subproblems $P_k$ for $k = k_{min},..., k_{max}$ appeared to be an interesting approach (\cite{Vasquez2001,Vasquez2005,Vimont2008}) particularly because the additional constraint ($1 \cdot x = k$) provides tighter upper bounds than the classical LP--relaxation. This approach takes also into consideration a constraint based on the reduced costs of the non--basic variables. Let us consider the upper bound $UB=c \cdot \bar x$, where $\bar x$ is the optimal solution of the LP--relaxation. Let $\bar{c}$ be the vector of the reduced costs and $N$ the indexes of the non--basic variables. If we know a lower bound $LB$ $\in \mathbb{N}$ of the problem, then each better solution $x$ must satisfy the following constraint:
\begin{eqnarray}
\label{CR}
\sum_{j \in N \mid \bar x_j=0} |\bar{c}_j| x_j +  \sum_{j \in N \mid \bar x_j=1} |\bar{c}_j| (1-x_j) \leq UB - LB \label{rc}
\end{eqnarray}

 (the reader is referred to \cite{Balas1980,Oliva2001} and \cite{Vimont2008} for more details on the subject). The use of the constraint (\ref{CR}), also called the \textsl{reduced costs constraint}, is twofold: (i) it enables us to  identify the unfeasibility (with respect to the constraint $c \cdot x \geq LB + 1$) of some partial solutions and (ii) it allows us to fix those variables $x_j$ with $|c_j| > UB - LB$ at their optimal value $\bar x_j$. Both the tight upper bound and the good lower bound enhance the efficiency of the reduced costs constraint.

To summarize, our approach consists of decomposing the search space in several hyperplanes then solving the LP-relaxation for each one in order to generate the reduced costs constraint. The exploration is then carried out partially and iteratively for each hyperplane with resolution search until all the search space is explored.


Indeed, thanks to the structure of resolution search, $P$ can be solved by progressively exploring each of the hyperplanes ($1 \cdot x = k,\ k \in \mathbb{N}$). Let \texttt{CFamily[$k$]} be the family of clauses associated to resolution search for the subproblem $P_k$. At each step of the search,  \texttt{CFamily[$k$]} provides all the information about the state of the search: the terminal nodes recorded at this stage and the next node to explore. It is possible to execute some iterations of resolution search at a given $P_k$, then continue to another $P_{k'}$ and go on with the subproblem $P_k$ again without any loss of information. The algorithm \ref{irs} detail the \texttt{Iterative\_RS} algorithm  which corresponds to resolution search limited to a given number of iterations and the algorithm \ref{hyp} shows the hyperplane exploration using \texttt{Iterative\_RS}. This way of exploration enforces the diversification and the convergence of the search. Note that, in algorithm \ref{hyp}, \texttt{greedy} is a simple greedy function used for providing a first lower bound.

\begin{figure}[h!]
   \begin{minipage}[b]{0.40\linewidth}
   \begin{footnotesize}
     \begin{tabular}{l}
\texttt{Iterative\_RS(Nb\_Iter, LB,$\mathscr{F}$)}\\
\texttt{\{}\\
\hspace{0.5cm}\texttt{iter = 0;}\\
\hspace{0.5cm}\texttt{while(iter < Nb\_Iter)} \{\\
\hspace{1cm}\texttt{try = obstacle(} u($\mathscr{F}$),\texttt{LB},$S$);\\
\hspace{1cm}\texttt{if(try > LB)}\hspace{0.3cm}\texttt{LB = try;}\\
\hspace{1cm}\texttt{add} $S$ \texttt{to} $\mathscr{F}$ and update $\mathscr{F}$;\\
\hspace{1cm}\texttt{if($(*, *, ..., *) \in \mathscr{F}$)}\hspace{0.3cm} \texttt{Break;} \\
\hspace{1cm}\texttt{iter++};\\
\hspace{0.5cm}\}\\
\}
\end{tabular}
\end{footnotesize}
      \caption{Iterative resolution search \label{irs}}
   \end{minipage}\hfill
   \begin{minipage}[b]{0.48\linewidth}
      \centering
        \begin{footnotesize}
    \begin{tabular}{l}
\texttt{LB = greedy();}\\
\texttt{Compute the bounds $k_{min}$ and $k_{max}$;}\\
\texttt{Set $\mathcal{K} $ = \{$k_{min}$, ..., $k_{max}$;\}}\\
\texttt{for($k=k_{min};k\leq k_{max};k++)$}\\
\hspace{0.5cm}\texttt{CFamily[$k$] $ = \emptyset$} \\
\texttt{While ($\mathcal{K} \neq \emptyset $) do} \{\\
\hspace{0.5cm}\texttt{Choose $k \in \mathcal{K} $;}\\
\hspace{0.5cm}\texttt{Choose Nb\_Iter\_k $\geq 1$;} \\
\hspace{0.5cm}\texttt{Iterative\_RS(Nb\_iter\_k,LB,CFamily[$k$]);} \\
\hspace{0.5cm}\texttt{if($(*, *, ..., *) \in $ CFamily[$k$])}\hspace{0.3cm} $\mathcal{K}  = \mathcal{K}  - \{k\}$;\\
\}\\
\end{tabular}
      \caption{Hyperplanes exploration \label{hyp}}
         \end{footnotesize}
   \end{minipage}
\end{figure}

\section{Resolution search and branch \& bound combination}
In this section, we detail the exploration function \texttt{obstacle} embedding  the branch \& bound algorithm.

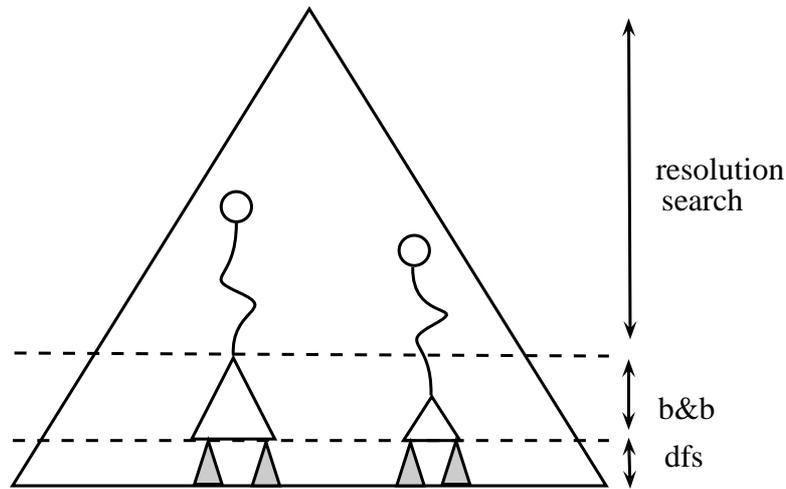
\begin{figure}[!h]
\begin{center}
\hspace{2cm}
\scalebox{1} 
{
\begin{pspicture}(0,-3.22)(10.378437,3.2)
\definecolor{color363b}{rgb}{0.8,0.8,0.8}
\pstriangle[linewidth=0.04,dimen=outer](3.94,-3.18)(7.88,6.38)
\psline[linewidth=0.04cm,linestyle=dashed,dash=0.16cm 0.16cm](0.04,-2.56)(7.88,-2.56)
\psline[linewidth=0.04cm,linestyle=dashed,dash=0.16cm 0.16cm](0.06,-1.4)(7.88,-1.44)
\pstriangle[linewidth=0.04,dimen=outer](2.94,-2.56)(1.16,1.14)
\pstriangle[linewidth=0.04,dimen=outer](5.55,-2.58)(0.82,0.64)
\pscircle[linewidth=0.04,dimen=outer](5.32,-0.04){0.22}
\psbezier[linewidth=0.04](2.98,0.32152376)(2.98,-0.4764442)(2.5,-0.36)(3.02,-0.6)(3.54,-0.84)(2.9,-0.8)(2.94,-1.4340059)
\psbezier[linewidth=0.04](5.3,-0.26)(5.28,-1.0)(6.08,-0.74)(5.58,-1.02)(5.08,-1.3)(5.56,-1.2)(5.54,-1.96)
\pstriangle[linewidth=0.04,dimen=outer,fillstyle=solid,fillcolor=color363b](2.61,-3.16)(0.42,0.66)
\pstriangle[linewidth=0.04,dimen=outer,fillstyle=solid,fillcolor=color363b](3.37,-3.18)(0.42,0.66)
\pstriangle[linewidth=0.04,dimen=outer,fillstyle=solid,fillcolor=color363b](5.27,-3.18)(0.42,0.66)
\pstriangle[linewidth=0.04,dimen=outer,fillstyle=solid,fillcolor=color363b](5.87,-3.16)(0.42,0.66)
\psline[linewidth=0.04cm,arrowsize=0.05291667cm 3.03,arrowlength=1.4,arrowinset=0.4]{<->}(8.16,-1.22)(8.14,3.04)
\psline[linewidth=0.04cm,arrowsize=0.05291667cm 3.0,arrowlength=1.4,arrowinset=0.4]{<->}(8.14,-2.42)(8.14,-1.48)
\psline[linewidth=0.04cm,arrowsize=0.05291667cm 3.0,arrowlength=1.4,arrowinset=0.4]{<->}(8.16,-3.2)(8.14,-2.52)
\usefont{T1}{ptm}{m}{n}
\rput(9.338125,1.05){resolution}
\usefont{T1}{ptm}{m}{n}
\rput(9.110156,0.63){search}
\usefont{T1}{ptm}{m}{n}
\rput(9.4425,-1.73){}
\usefont{T1}{ptm}{m}{n}
\rput(8.9,-2.13){b\&b}
\usefont{T1}{ptm}{m}{n}
\rput(8.867812,-2.77){dfs}
\pscircle[linewidth=0.04,dimen=outer](2.98,0.54){0.22}
\end{pspicture}
}
\caption{Global view of the exploration process of an hyperplane}\label{schema1}
\end{center}
\end{figure}

Starting with the node $u(\mathscr{F})$ given by the path--like family $\mathscr{F}$, \texttt{obstacle} replaces step--by--step the components $\ast$ of  $u(\mathscr{F})$ by 0 or 1 which constructs the node $u^+$. If $u^+$ is a terminal node, the function provides a minimal clause $S$ such that $S \sqsubseteq u^*$ and $S$ is a terminal node. Our implementation of \texttt{obstacle} is based on the reduced--costs constraint (\ref{rc}) presented in section \ref{genapp}. At the beginning of the procedure, the LP--relaxation of the problem is solved for each available hyperplane in order to give us the information needed for the reduced costs constraint. Then \texttt{obstacle} proceeds to the following steps:

\begin{itemize}

\item The first step, called \textsl{consistency phase}, consists of checking the feasibility of $u(\mathscr{F})$. Initially, we define a value $gap = UB - LB$. If a constraint is violated, the descent phase stops and the corresponding partial instantiation is recorded as the clause $S$ in $\mathscr{F}$. At the same time, the reduced costs constraint (\ref{rc}) is checked and the $gap$ value is updated: for each non--basic variable set at the opposite of its optimal value ($1 - \bar x_j$), its reduced costs ($\bar c_j$) is subtracted from the gap. If it happens that $gap < 0$ then the current partial solution is a terminal node. In this case, $S$ is only composed of the variables set at the opposite of their optimal value in $u(\mathscr{F})$.

\item If $u(\mathscr{F})$ is feasible, we go to the next step which we call \textsl{implicit waning phase}. This phase consists of branching on all the remaining free variables with a reduced cost greater than $gap$. Those variables must be fixed at their optimal value for satisfying the reduced costs constraint (\ref{rc}). The branching decisions taken in this phase are just a consequence of the instantiated variables in $u(\mathscr{F})$. Consequently they can be removed from $S$.

\item Then the algorithm starts the so--called \textsl{waxing phase} which consists of assigning values to free variables. The chosen strategy is to select the free variable with the greater absolute reduced cost value and to assign its optimal value $\bar x_j$ to it. Obviously, each time a branching is done, the feasibility of the current partial solution is verified and in case of fail, the waxing phase stops and the corresponding clause $S$ is added to $\mathscr{F}$. Note that the variables set in the implicit waning phase are still not taken into account in $S$.

\item Once the number of remaining free variables is less than or equal to a given number \texttt{spb\_size}, the waxing phase stops and the corresponding subproblem is solved with a branch \& bound algorithm. This subproblem includes the free variables with the lowest reduced cost and the basic variables. Obviously, since the branch \& bound algorithm explores the whole subtree corresponding to these variables, the clause $S$ does not contain any branching choices made during this phase. Only the branching decisions taken during the consistency phase \textsl{and/or} the waxing phase are considered.

\item The algorithm used to enumerate the variables with the lowest reduced cost, and the basic variables, is widely inspired by a previous one published by \cite{Vimont2008}. As it is represented in figure \ref{schema1} (which summarizes the exploration process), this algorithm embeds a specific Depth First Search (dfs) procedure for solving small subproblems with the last $20$ variables.
\end{itemize}
\vspace{-0.7cm}
\vspace{-1.2cm}
\section{Computational results}
Our algorithm has been experimented on the well known OR-Library set of 01MKP instances proposed by \cite{Beasley1990}. Each instance of this set is denoted cb$m$.$n$\_$r$ where $m$ is the number of constraints, $n$ the number of variables and $r$ the instance's number. The 01MKP OR-Library  benchmark is composed of $5$, $10$ and $30$  constraint instances and $100$, $250$ and $500$ variables instances. There are $9$ set of $30$ instances with each $n \times m$ combination.

We obtained better proof time than the exact published approaches of \cite{James2005}, \cite{Vimont2008} and the commercial software CPLEX 9.2, for the $10$ constraint, $250$ variable instances and the $5$ constraint, $500$ variable instances. We observed also that for the $5$ constraint, $500$ variable instances, the times required for obtaining the best solutions are clearly better than the ones provided by the  best known heuristics on these instances (\cite{Vasquez2005} and \cite{Wilbaut2008}). Moreover, our algorithm proved the optimality of all the cb10.500 instances. The corresponding optimal values, which were previously unknown, are exposed in table \ref{results}. The column $z^{opt}$ is the optimal value of the instance, \textsl{opt. (h)} is the time in hours required for obtaining the optimal value and \textsl{proof (h)} is the time in hours for proving the optimality of the value  $z^{opt}$. The column  $z^{opt}-\underline{z}$ corresponds to the gap between the optimal value and the previously best known solution: \textsl{(vv)} indicates that $\underline{z}$ was first found by \cite{Vasquez2005} and \textsl{(wh)} indicates that $\underline{z}$ was first found by \cite{Wilbaut2008}.

\begin{table}[htb]
\begin{scriptsize}
\centerline{
\begin{tabular}{|llrrr||llrrr|}
\hline
Instance & $z^{opt}$ & opt. (h) & proof (h) & $z^{opt}-\underline{z}$&Instance & $z^{opt}$ & opt. (s) & proof (h) & $z^{opt}-\underline{z}$\\
\hline
cb10.500\_0&117821&24,5&567,2&\textbf{+10}(vv)&cb10.500\_15&215086&0&43,9&0\\
cb10.500\_1&119249&68,4&272,9&\textbf{+17}(vv)&cb10.500\_16&217940&13,4&36,1&0\\
cb10.500\_2&119215&18,6&768,3&0&cb10.500\_17&219990&150,8&348,8&0\\
cb10.500\_3&118829&47,4&89,6&\textbf{+4}l(wh)&cb10.500\_18&214382&12,7&57,8&\textbf{+7}(vv)\\
cb10.500\_4&116530&86,1&2530,3&\textbf{+16}(wh)&cb10.500\_19&220899&0,2&21,3&0\\
cb10.500\_5&119504&2,3&188&0&cb10.500\_20&304387&6,6&8,2&0\\
cb10.500\_6&119827&2,7&128&0&cb10.500\_21&302379&0&8,4&0\\
cb10.500\_7&118344&161,7&179,6&\textbf{+11}(wh)&cb10.500\_22&302417&67,2&105,5&\textbf{+1}(vv)\\
cb10.500\_8&117815&86,3&219,9&0&cb10.500\_23&300784&0,9&3,8&0\\
cb10.500\_9&119251&3,1&354,9&0&cb10.500\_24&304374&0,1&16,8&0\\
cb10.500\_10&217377&0&515,8&0&cb10.500\_25&301836&29,7&30,9&0\\
cb10.500\_11&219077&0,5&437,6&0&cb10.500\_26&304952&0&18,5&0\\
cb10.500\_12&217847&0&5,5&0&cb10.500\_27&296478&1,1&9,3&0\\
cb10.500\_13&216868&0&104,4&0&cb10.500\_28&301359&8,1&39,1&0\\
cb10.500\_14&213873&59,4&1382,1&\textbf{+14}(vv)&cb10.500\_29&307089&1,2&4,4&0\\
\hline
\end{tabular}}
\end{scriptsize}
\caption{Results obtained on the $10$ constraint, $500$ variable instance of the  \texttt{OR-Library} \label{results}}
\end{table}
\section{Conclusion}

Although our implementation of resolution search is quite far from the original one proposed by Chv\'atal, especially because we have exploited specific structures of the 01MKP (hybridization with branch \& bound, $1.x=k$ hyperplane decomposition, reduced cost constraint to generate implicitly the partial instantiations responsible for the fails, etc.), we showed that resolution search is a promising framework for designing efficient algorithms. Since the proof times are long for the $10$ constraint, $500$ variable instances, we plan to improve again our algorithm in order to accelerate the resolution process in the hope of maybe closing the OR-Library 01MKP benchmark by solving the 60 last 30 constraint 250/500 variable instances.

\end{document}